\documentclass[sigconf, nonacm]{acmart}
\usepackage{subcaption}
\usepackage{listings}  
\usepackage{url}
\usepackage{tikz}
\usetikzlibrary{shapes,arrows,positioning}
\usepackage{microtype}

\AtBeginDocument{%
}

\setcopyright{acmlicensed}
\copyrightyear{2025}
\acmYear{2025}
\acmDOI{XXXXXXX.XXXXXXX}

\acmConference[Conference acronym 'XX]{Make sure to enter the correct
conference title from your rights confirmation email}{June 03--05,
2025}{Woodstock, NY}

\acmISBN{978-1-4503-XXXX-X/2025/06}

\newcommand{\ruleforge}{\texttt{RuleForge}}  

\usepackage{xcolor}
\lstdefinelanguage{json}{
string=[b]{"},
stringstyle=\color{blue},
comment=[l]{:},
commentstyle=\color{black},
morecomment=[l]{,}
}

\begin{document}

\title{\ruleforge: Automated Generation and Validation for Web Vulnerability Detection at Scale}


\author{Ayush Garg}
\email{sblayush@gmail.com}
\affiliation{%
  \country{USA}
}

\author{Sophia Hager}
\orcid{0009-0000-8470-8900}
\email{shager2@jh.edu}
\affiliation{%
  \institution{Johns Hopkins University}
  \country{USA}
}

\author{Jacob Montiel}
\orcid{0000-0003-2245-0718}
\email{jacobmlx@amazon.com}
\affiliation{%
  \institution{Amazon Web Services}
  \country{USA}
}

\author{Aditya Tiwari}
\email{aditiwar@amazon.com}
\affiliation{%
  \institution{Amazon Web Services}
  \country{USA}
}

\author{Michael Gentile}
\email{mgent@amazon.com}
\affiliation{%
  \institution{Amazon Web Services}
  \country{USA}
}

\author{Zach Reavis}
\email{hackary@amazon.com}
\affiliation{%
  \institution{Amazon Web Services}
  \country{USA}
}

\author{David Magnotti}
\email{dmagnott@amazon.com}
\affiliation{%
  \institution{Amazon Web Services}
  \country{USA}
}

\author{Wayne Fullen}
\orcid{0009-0001-7666-6744}
\email{wfullen@amazon.com}
\affiliation{%
  \institution{Amazon Web Services}
  \country{USA}
}

\renewcommand{\shortauthors}{A. Garg et at.}

\begin{abstract}
Security teams face a challenge: the volume of newly disclosed Common Vulnerabilities and Exposures (CVEs) far exceeds the capacity to manually develop detection mechanisms. In 2025, the National Vulnerability Database published over 48,000 new
vulnerabilities~\cite{nvd2025dashboard}, motivating the need for automation. We present \ruleforge, an AWS internal system that automatically generates detection rules—JSON-based patterns that identify malicious HTTP requests exploiting specific vulnerabilities—from structured Nuclei templates describing CVE details. Nuclei templates provide standardized, YAML-based vulnerability descriptions that serve as the structured input for our rule generation process.

This paper focuses on \ruleforge's architecture and operational deployment for CVE-related threat detection, with particular emphasis on our novel LLM-as-a-judge (Large Language Model as judge) confidence validation system and systematic feedback integration mechanism. This validation approach evaluates candidate rules across two dimensions---sensitivity (avoiding false negatives) and specificity (avoiding false positives)---achieving AUROC of 0.75 and reducing false positives by 67\% compared to
synthetic-test-only validation in production. Our 5×5 generation strategy (five parallel candidates with up to five refinement attempts each) combined with continuous feedback loops enables systematic quality improvement. We also present extensions enabling rule generation from unstructured data sources and demonstrate a proof-of-concept agentic workflow for multi-event-type detection. Our lessons learned highlight critical considerations for applying LLMs to cybersecurity tasks, including overconfidence mitigation and the importance of domain expertise in both prompt design and quality review of generated rules through human-in-the-loop validation.
\end{abstract}



\maketitle

\section{INTRODUCTION}

In today's rapidly evolving threat landscape, cybersecurity teams face an escalating challenge: the volume of newly discovered vulnerabilities far outpaces the capacity to develop detection mechanisms. CVEs are publicly disclosed security flaws that attackers can exploit to compromise systems, steal data, or disrupt services. The National Vulnerability Database (NVD) continuously publishes new high-severity vulnerabilities, creating an urgent need for automated approaches to threat detection. Security researchers and practitioners have developed tools like Nuclei~\cite{Nuclei}—an open-source vulnerability scanner that uses structured templates to describe exploit patterns—to help identify and validate vulnerabilities. However, converting these vulnerability descriptions into production-ready detection rules that can operate at cloud scale remains a bottleneck, presenting both a critical security challenge and a research opportunity.

\ruleforge~is a system designed at AWS to automatically generate rules for detecting and mitigating malicious requests exploiting CVEs. These rules are used by a security component that normalizes heterogeneous event log data into Open Cybersecurity Schema Framework (OCSF) format and uses JSON-based rules to classify and tag security events, matching potential exploit attempts to specific CVEs. Currently, detection rules are manually created and validated, a time-intensive process. In 2025, the National Vulnerability Database (NVD) published over 15 times more high-severity vulnerabilities than an internal team could manually create detection rules for, demonstrating that manual rule creation cannot keep pace with the growing threat landscape. This gap between vulnerability disclosure and rule creation presents a research opportunity to develop automated approaches that can strengthen security postures and improve threat detection coverage across cloud infrastructure.

\ruleforge~uses a series of prompts with generative AI (via Amazon Bedrock) to create rules directly from Nuclei templates, which offers the opportunity to speed up rule creation and close the gap of CVEs in the backlog. However, the dependency on Nuclei templates introduces two limitations: (1) the template format lacks expressiveness for certain complex vulnerability types, and (2) only a subset of CVEs are converted to Nuclei templates by the security research community, constraining the system's coverage.

The rules should have high recall (should not falsely flag malicious rules as benign), and high precision (should not falsely flag benign requests as malicious). Precision is the priority, as false positives would cause ordinary users to be marked as bad actors. Evaluation presents a challenge: generally evaluation uses labeled data, in this case malicious requests and benign requests with similarities to malicious requests. While there may be examples of malicious requests in the description of the vulnerability, there are no similar benign requests, and so the rule will not be validated on challenging edge cases.

While \ruleforge~demonstrated significant productivity improvements, two key research challenges emerged that required investigation to enhance the system's capabilities and reliability:

\begin{itemize}
\item \textbf{Additional validation of candidate rules:} The existing validations provide useful information about whether a rule should be accepted, but may be unreliable (e.g. incorrectly labeled synthetic tests) or have limited applicability. Additionally, neither validation provides much specific information about what part of the rule fails and why.
\item \textbf{Dependency on Nuclei templates:} The current system only generates rules based on the structured data from Nuclei, and has not been evaluated with unstructured data (e.g. blog posts, NVD descriptions of vulnerabilities) where there may be variability in the type and presentation of information. Running on unstructured data would expand the amount of rules that can be created.
\end{itemize}

This paper presents the \ruleforge~system architecture, design principles, and operational workflow, with particular focus on two major enhancements developed to address these challenges: an LLM-as-a-judge confidence validation mechanism and an extension to support unstructured data sources. Section~\ref{sec:ruleforge_background} provides detailed background on the \ruleforge~system architecture, Section~\ref{sec:rule_select} explains how the CVEs are being selected, Section~\ref{sec:validation} describes the comprehensive candidate rule validation pipeline including the LLM-as-a-judge validation system, Section~\ref{sec:production} presents production deployment results, Section~\ref{sec:extensions} discusses \ruleforge~extensions exploring generalization and adaptability (including unstructured data generation and an agentic workflow proof-of-concept), Section~\ref{sec:lessons_learned} presents lessons learned, Section~\ref{sec:related_work} covers related work, and Section~\ref{sec:conclusion} provides our conclusions and future work directions.

\section{THE RULEFORGE SYSTEM}\label{sec:ruleforge_background}

\ruleforge~is an automated system designed to generate detection rules for CVEs at scale. These rules are used by a security component that normalizes heterogeneous event log data into Open Cybersecurity Schema Framework (OCSF) format and uses JSON-based rules to classify and tag security events, matching potential exploit attempts to specific CVEs.

\subsection{System Architecture}

\begin{figure*}[ht!]
\centering
\includegraphics[width=\linewidth]{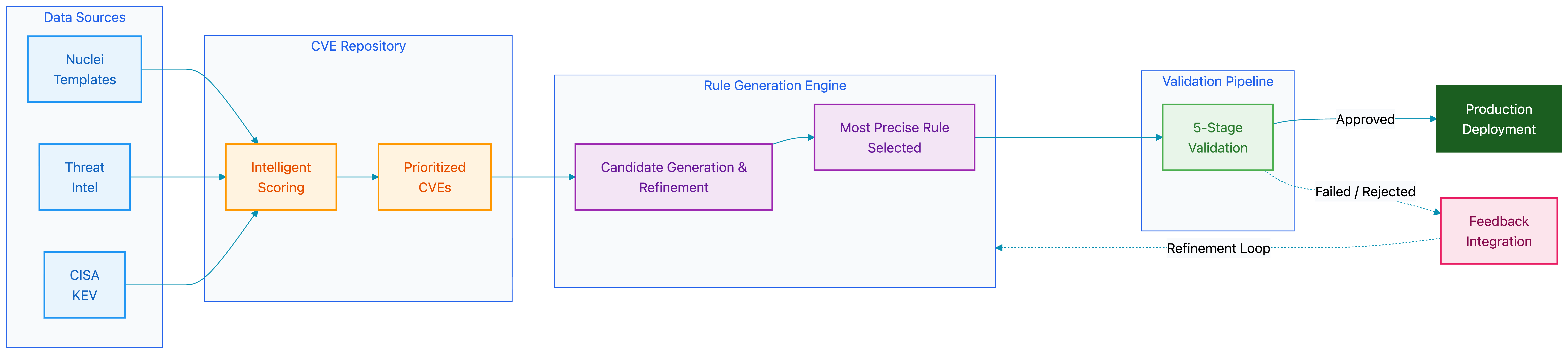}
\caption{\ruleforge~architecture showing CVE Repository, Rule Generation Engine, Validation Pipeline, and Feedback Integration components.}
\label{fig:architecture}
\end{figure*}

The \ruleforge~system consists of four main components, depicted in Figure~\ref{fig:architecture}:

\textbf{CVE Repository.} A Lambda function pulls CVE data weekly from Nuclei templates, applies intelligent scoring based on content analysis and threat intelligence (CISA KEV, news feeds), and stores prioritized metadata in DynamoDB with full templates in S3, providing ranked structured input for rule generation.

\textbf{Rule Generation Engine.} Runs on AWS Fargate with Amazon Bedrock, using the DSPy framework~\cite{khattab2024dspy} for LLM interactions. For each CVE, the system generates five parallel rule candidates with different configurations at work, increasing the likelihood of producing high-quality detection rules by exploring diverse approaches simultaneously. Each candidate can undergo up to five refinement attempts based on validation feedback, with systemic feedback from validation failures fed back to the LLM for iterative improvement.

\textbf{Validation Pipeline.} A comprehensive five-stage process ensures production-ready quality: (1) synthetic testing with seven malicious and three benign HTTP requests, requiring correct identification of at least eight out of ten; (2) confidence scoring using an LLM-as-a-judge to calculate Sensitivity and Specificity scores; (3) comparison against five billion records of production web traffic data, requiring 10--500 matches; (4) IP reputation validation against AWS internal sources and third-party threat intelligence; and (5) human review by security engineers via internal code review tools.

\textbf{Feedback Integration.} The system captures both systemic feedback when rules fail validation stages and human feedback from security engineer code reviews, feeding these insights back into the rule generation process to progressively improve rule quality over time.

\subsection{Key Design Decisions}

\begin{figure}[!htbp]
\centering
\includegraphics[width=0.95\columnwidth]{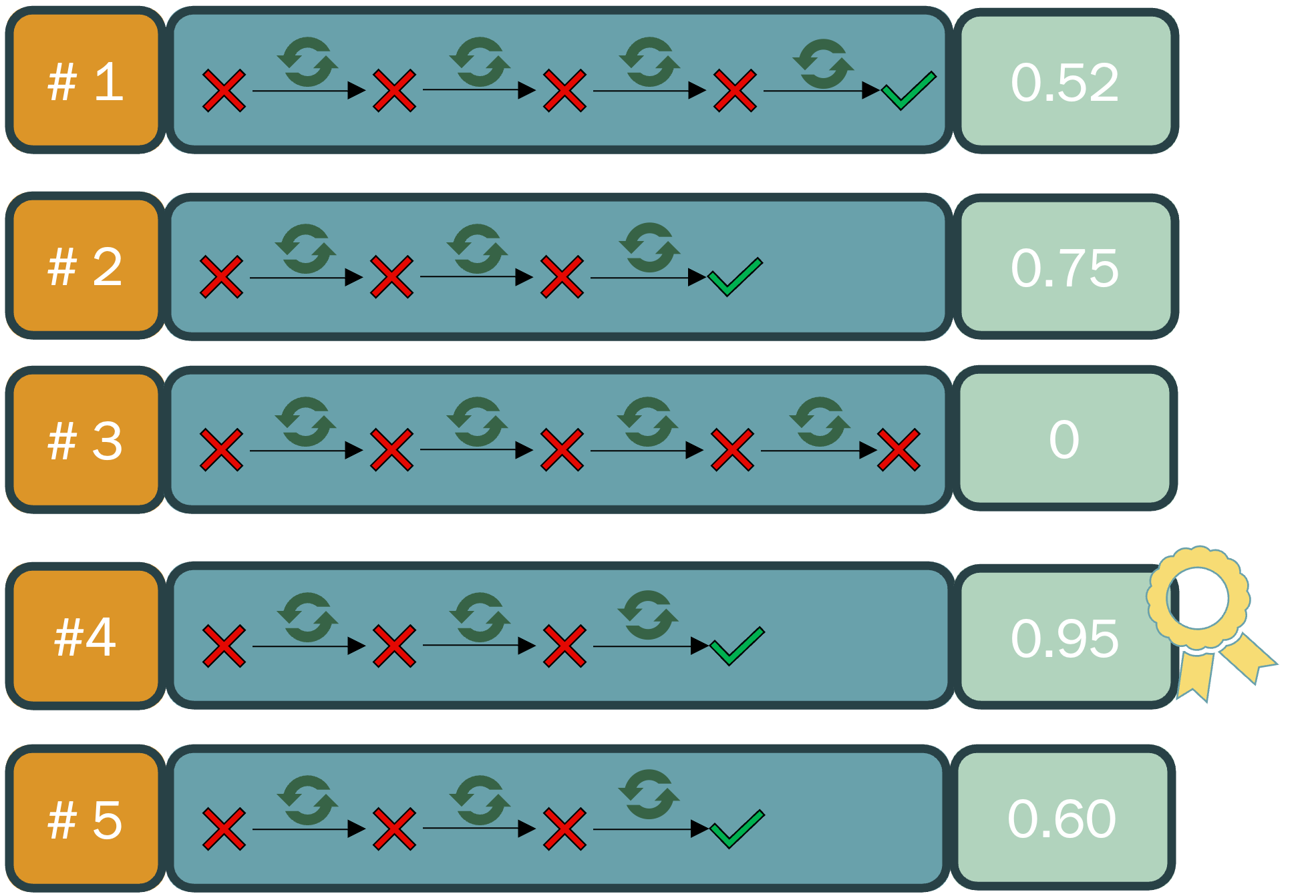}
\caption{5×5 Generation Strategy showing five parallel rule candidates with confidence scores and iterative refinement process. The system generates multiple candidates simultaneously and selects the best performer based on validation results.}
\label{fig:five_by_five_strategy}
\end{figure}

Three core design principles guide \ruleforge's architecture:

\textbf{Parallel Candidate Generation with Systematic Feedback.} Rather than generating a single rule per CVE, our 5×5 strategy creates five parallel candidates simultaneously with different LLM temperature settings (randomly sampled between 0.7 and 0.9), each capable of up to five refinement attempts based on validation feedback, see~Figure~\ref{fig:five_by_five_strategy}. This approach allows the system to explore various detection approaches and converge on the most effective logic through iterative refinement, increasing the probability of generating at least one high-quality rule per CVE.

\textbf{Multi-Stage Validation.} The comprehensive validation pipeline ensures that generated rules meet production quality standards before deployment, balancing automated testing with human expertise to minimize false positives while maintaining high detection rates.

\begin{figure}[htbp]
\centering
\includegraphics[width=0.95\columnwidth]{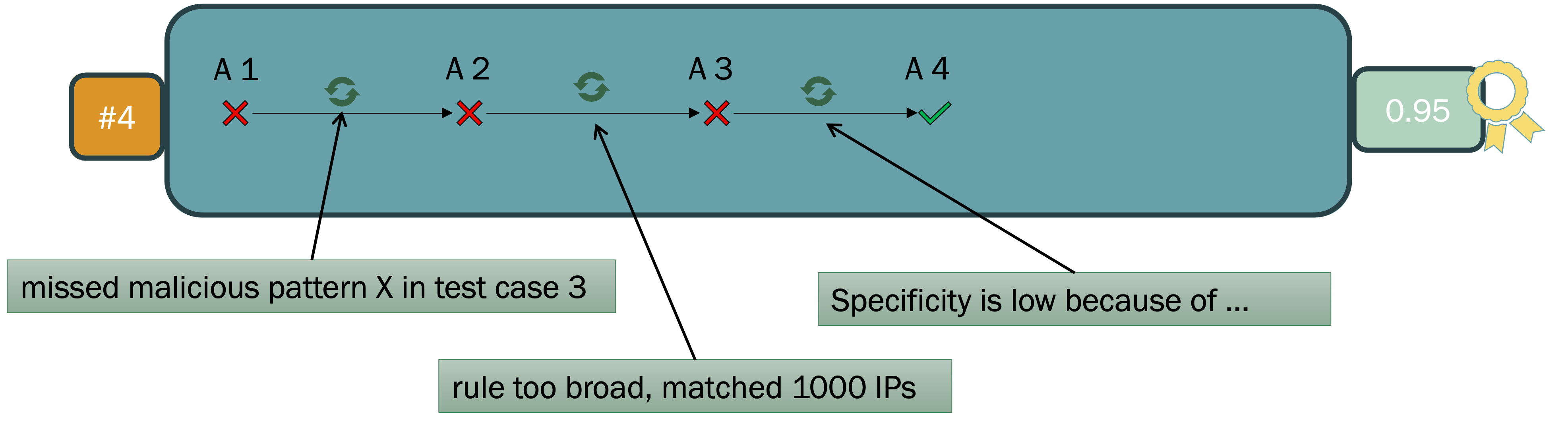}
\caption{Systemic Feedback Loop mechanism showing how validation failures generate specific feedback that is integrated into subsequent rule generation attempts. Examples include feedback about missed patterns, overly broad rules, and specificity issues.}
\label{fig:feedback_loop}
\end{figure}

\textbf{Continuous Learning with Systematic Feedback Integration.} By capturing and integrating feedback from both automated validation failures and human reviews, the system creates a comprehensive learning loop that progressively improves rule quality and adapts to evolving threat patterns, see~Figure~\ref{fig:feedback_loop}. This feedback mechanism enables the system to learn from failures and systematically improve its generation capabilities over time.

\section{CVE Rule Selection for Generation}\label{sec:rule_select}

The CVE Rule Selection component functions as a critical filtering mechanism that determines which vulnerabilities warrant automated rule generation within the \ruleforge~pipeline. This component implements a multi-criteria weighted scoring algorithm that evaluates Common Vulnerabilities and Exposures (CVEs) across three primary dimensions: content-based analysis, threat intelligence integration, and organizational relevance. Content-based analysis employs keyword matching against CVE descriptions to identify high-impact vulnerability characteristics, with elevated priority assigned to vulnerabilities affecting enterprise infrastructure vendors, remote code executions that use command injection or SQL injection in enterprise applications, and authentication bypass mechanisms. Conversely, the system applies negative weighting to reduce noise from low-impact vulnerability categories, such as isolated plugin issues, thereby focusing computational resources on threats with broader organizational impact.

\begin{figure}[htbp]
\centering
\begin{tikzpicture}[node distance=3.5cm,auto,
block/.style={rectangle, draw, fill=blue!20, text width=2cm, text centered, rounded corners, minimum height=1cm},
decision/.style={diamond, draw, fill=yellow!20, text width=1.5cm, text centered, minimum height=1cm},
arrow/.style={thick,->,>=stealth}]

\node [block] (input) {CVE Input\\(No Existing Rules)};
\node [block, below left of=input] (content) {Content Analysis};
\node [block, below of=input] (cisa) {CISA KEV Check};
\node [block, below right of=input] (news) {News Feed Check};
\node [block] (score) at ([yshift=-1.5cm]cisa.south) {Calculate Weighted Score};
\node [block] (rank) at ([yshift=-1cm]score.south) {Rank CVEs by Priority};
\node [block] (generate) at ([yshift=-1cm]rank.south) {Generate Rules\\(Highest Priority First)};

\draw [arrow] (input) -- (content);
\draw [arrow] (input) -- (cisa);
\draw [arrow] (input) -- (news);
\draw [arrow] (content) -- (score);
\draw [arrow] (cisa) -- (score);
\draw [arrow] (news) -- (score);
\draw [arrow] (score) -- (rank);
\draw [arrow] (rank) -- (generate);

\end{tikzpicture}
\caption{CVE Rule Selection Process Flow. The system processes CVEs without existing rules through three parallel analysis components: content-based keyword matching, CISA KEV status verification, and cybersecurity news feed presence checking. These inputs feed into a weighted scoring algorithm that ranks CVEs by priority, with the highest-scoring vulnerabilities selected first for automated rule generation.}
\label{fig:cve-selection}
\end{figure}
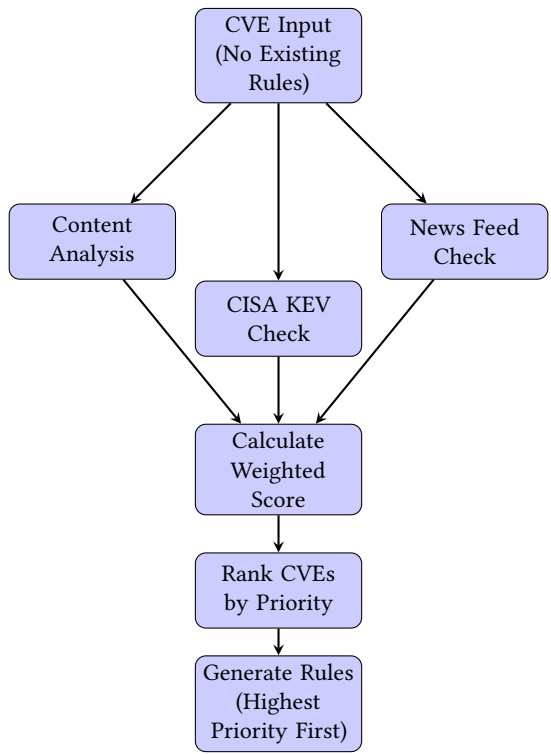

The selection algorithm incorporates real-time threat intelligence from authoritative sources including the CISA Known Exploited Vulnerabilities (KEV) catalog and cybersecurity news feeds, substantially increasing priority scores for CVEs with confirmed active exploitation or emerging threat status. Following weight calculation across these criteria, the component applies configurable threshold filtering to select CVEs for downstream rule generation, while maintaining comprehensive audit trails documenting selection decisions and scoring rationale. This intelligent filtering approach enables the \ruleforge~service to optimize both security coverage and operational efficiency by concentrating rule generation efforts on vulnerabilities that present the greatest risk to enterprise environments, as illustrated in Figure~\ref{fig:cve-selection}.

\section{Candidate Rule Validation}\label{sec:validation}

\begin{figure}[ht!]
\centering
\includegraphics[width=\columnwidth]{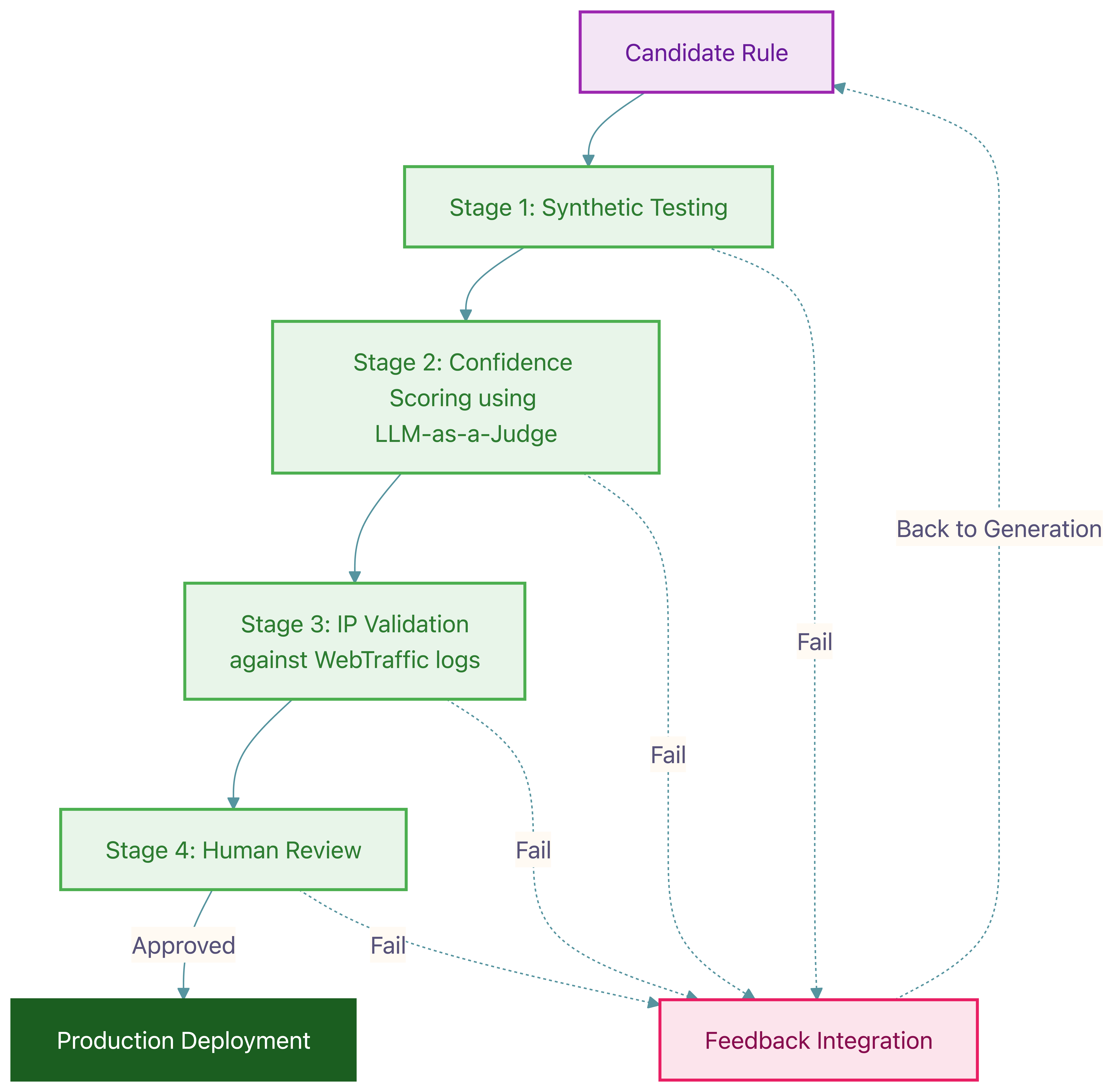}
\caption{Candidate Rule Validation Pipeline. Each generated rule passes sequentially through four stages: (1) Synthetic Testing, (2) LLM-as-a-Judge Confidence Scoring, (3) IP Validation against production web traffic, and (4) Human Review. Approved rules proceed to Production Deployment. Failures at any stage are routed to Feedback Integration, which provides stage-specific reasoning back to the Rule Generation Engine for iterative refinement.}
\label{fig:validation_stages_fig}
\end{figure}

\ruleforge~employs a multi-stage validation pipeline, illustrated in Figure~\ref{fig:validation_stages_fig}, to ensure that generated rules meet production quality standards before deployment. This section describes the core validation mechanisms and introduces the LLM-as-a-judge confidence validation system that enhances these baseline methods. Confidence validation implementation details are provided in Appendix~\ref{sec:confidence_prompts}.

\subsection{Synthetic Testing}

\ruleforge~prompts the LLM to generate ten synthetic test cases to validate each rule: seven malicious requests and three benign requests based on the examples of malicious requests contained in the CVE. If the candidate rule identifies malicious requests without flagging benign requests, it moves on to the next stage of evaluation; otherwise, the system retries rule generation. One limitation of this approach is that the synthetic nature of the test cases means the generated tests may be incorrect. There is therefore a two-error allowance for this test: rules are considered to pass the synthetic test validation even if they misclassify at most two test cases out of the ten. This tolerance accounts for potential errors in the synthetic test generation itself, where the LLM may occasionally mislabel a benign request as malicious or vice versa. For example, a rule passes if it correctly identifies all seven malicious requests and two of the three benign requests, or if it correctly identifies six of the seven malicious requests and all three benign requests.

While synthetic testing provides useful pass/fail signals, it offers limited
information about what specific part of a rule fails and why. This motivated
the development of an LLM-as-a-judge confidence validation system that provides
both a discriminative quality score and interpretable reasoning about rule
deficiencies.

\subsection{LLM-as-a-Judge Confidence Validation}\label{sec:confidence}

To construct an additional validation for the rules, we created an LLM-as-a-judge~\cite{zheng2023judging} system to answer questions about the rule and CVE pair. We construct the questions based on discussion over what human evaluators look for in rules:

\begin{itemize}
\item What is the probability that the rule described fails to flag some malicious requests described in the CVE?
\item What is the probability that the rule described is targeting a feature that correlates with the vulnerability rather than the vulnerability itself?
\end{itemize}

These are materially equivalent to "false negative" and "false positive", but provide more information about what to look for. The confidence evaluation explicitly prompts Claude 3.7 Sonnet to answer these questions, which returns two scores describing the model's estimate of the probability. Taking the complement of each results in a \textit{specificity score} (describing the probability of avoiding false positives) and a \textit{sensitivity score} (describing the probability of avoiding false negatives). This also returns two sets of \textit{reasoning} detailing the model's rationale for assigning the score.

\subsubsection{Score Evaluation}

To examine the effectiveness of the score, we multiply the sensitivity and specificity scores to create an overall confidence score. There are two metrics we use to evaluate the confidence score. To examine the calibration (i.e., how well the score corresponds to the actual error rate), we use Expected Calibration Error (ECE)~\cite{guo2017calibration}, where \textit{lower} ECE corresponds to better calibration. We quantify the score's discriminative power (i.e., whether a high-scoring prediction is more likely to be correct than a low-scoring prediction) using Area Under the Receiver Operating Characteristic curve (AUROC). This metric describes the probability that a randomly selected good rule is scored higher than a randomly selected incorrect rule, so \textit{higher} AUROC corresponds to a better quality score, with AUROC of 0.5 indicating random chance.

\subsubsection{Reasoning Analysis}

We manually compared the reasoning chains for nine rules to subject expert reasoning for the same rules. The small sample size was chosen due to the human time required for evaluation; further evaluation may be informative, but these rules were enough to demonstrate some trends. Overall, we found that the generated reasoning can resemble human reasoning: for instance, for a rule where a human evaluator noted "that SQL injection regex is too loose..." the LLM reasoning included that "...the regex pattern \texttt{$([{\wedge}=]+)=([{\wedge}']*')$} will catch any query parameter with a single quote, which is broader than just the specific vulnerability...".

The LLM reasoning matched human reasoning in six out of nine rules evaluated.
In the remaining three, the LLM was overly confident, missing limitations
identified by human reviewers. We note that this sample size is too small to
draw statistical conclusions---a 6/9 agreement rate is not distinguishable from
chance (binomial test $p = 0.25$ against a 50\% baseline)---and these results
should be interpreted as preliminary. Nevertheless, the qualitative alignment
between LLM and human reasoning in the agreeing cases suggests that confidence
reasoning is a promising source of actionable feedback, warranting further
evaluation on a larger sample.

\begin{figure}[htbp]
\centering
\includegraphics[width=0.95\columnwidth]{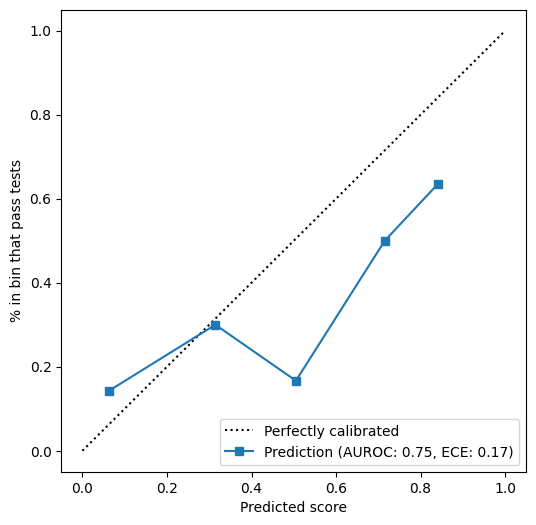}
\caption{Reliability diagram showing predicted confidence versus actual accuracy for the LLM-as-a-judge confidence scoring system. Points above the diagonal line indicate underconfidence, while points below indicate overconfidence. The system achieves AUROC of 0.75 and ECE of 0.17 on 100 test examples.}
\label{fig:reliability_diagram}
\end{figure}

On 100 generated examples, with IP validation~(Section~\ref{sec:ip_validation})
used to evaluate correctness, the LLM-as-a-judge system achieves AUROC of 0.75
and ECE of 0.17, showing that the score has relatively good discriminative power
and calibration (Figure~\ref{fig:reliability_diagram}). A limitation of this
approach is that the same model family (Claude) is used for both rule generation
and confidence evaluation, which may introduce self-preference
bias~\cite{panickssery2024llm}; future work could mitigate this by using a
different model family for evaluation.

\subsubsection{Integration into \ruleforge}

In early August 2025, the confidence scoring and reasoning system was built into \ruleforge. When a generated rule fails to meet the established thresholds (0.5 for sensitivity and 0.7 for specificity), the reasoning for the particular score is added to the feedback and passed as part of the prompt for another iteration of generation. This feedback mechanism enables iterative refinement, where the LLM receives specific guidance about why a rule was rejected and what aspects need improvement.

To evaluate the practical impact of this integration, we conducted a comparative experiment using fifty CVEs. Rules were generated twice using \ruleforge~with a budget of five retries per rule. In the baseline run, rules were evaluated only using synthetic tests to
determine acceptance or rejection, with no feedback provided beyond the number
of tests passed. In the enhanced run, the specificity and sensitivity scores were calculated for each candidate rule, and reasoning was appended to the prompt as feedback when scores fell below thresholds.

The results demonstrated substantial improvements in rule quality. The confidence scoring and feedback mechanism generated the same number of true positive rules as the baseline while achieving a 67\% decrease in false positives and a 71\% decrease in rules matching no IPs. This validates that the LLM-as-a-judge system effectively filters out poorly-constructed rules during generation without sacrificing detection capability, thereby reducing the burden of manual review for security engineers.

\subsection{IP Validation}\label{sec:ip_validation}

To validate rule quality against real-world traffic, we test generated rules
against production web traffic data (capped at 5 billion randomly sampled
records per month). This validation applies to rules that do not check request
bodies, where false positive risk is highest.

For each candidate rule, we measure the volume of matching requests in
production traffic. Rules matching 10--500 IPs proceed to IP enrichment, where
the associated IP addresses are cross-referenced with threat intelligence
sources including
MadPot~\cite{Ryland2023MadPot}\footnote{IP reputation is assessed using
MadPot and additional internal threat intelligence sources not described here.}.
A rule passes if more than 70\% of matched IPs are labeled as malicious;
otherwise, it fails.

\begin{figure}[htbp]
\centering
\includegraphics[width=0.95\columnwidth]{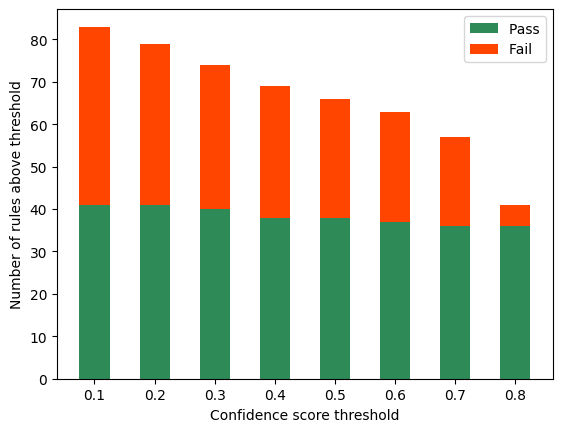}
\caption{Rules passing and failing IP validation across confidence score
thresholds. Higher thresholds preserve most passing rules while reducing
failures.}
\label{fig:confidence_threshold}
\end{figure}

Because this validation is computationally expensive, the confidence score
(Section~\ref{sec:confidence}) serves as a preliminary filter. As
Figure~\ref{fig:confidence_threshold} shows, increasing the confidence threshold
substantially reduces the number of rules failing IP validation while preserving
those that pass, confirming that confidence scoring effectively filters
low-quality rules before expensive IP enrichment.

\subsection{Human Validation}

The final stage of the validation pipeline involves human review by security engineers through Amazon's internal code review system. Each generated rule that passes the automated validation stages is submitted as an individual code review to the internal security team for expert evaluation. Security engineers assess the rule's technical correctness, potential for false positives, alignment with threat intelligence, and overall effectiveness in detecting the target vulnerability. This human validation ensures that domain expertise guides the final acceptance decision, maintaining the high quality standards required for production security systems while providing valuable feedback that is integrated back into the rule generation process for continuous improvement.

\section{Operational Evaluation}\label{sec:production}

\begin{figure}[htbp!]
\centering
\includegraphics[width=0.95\columnwidth]{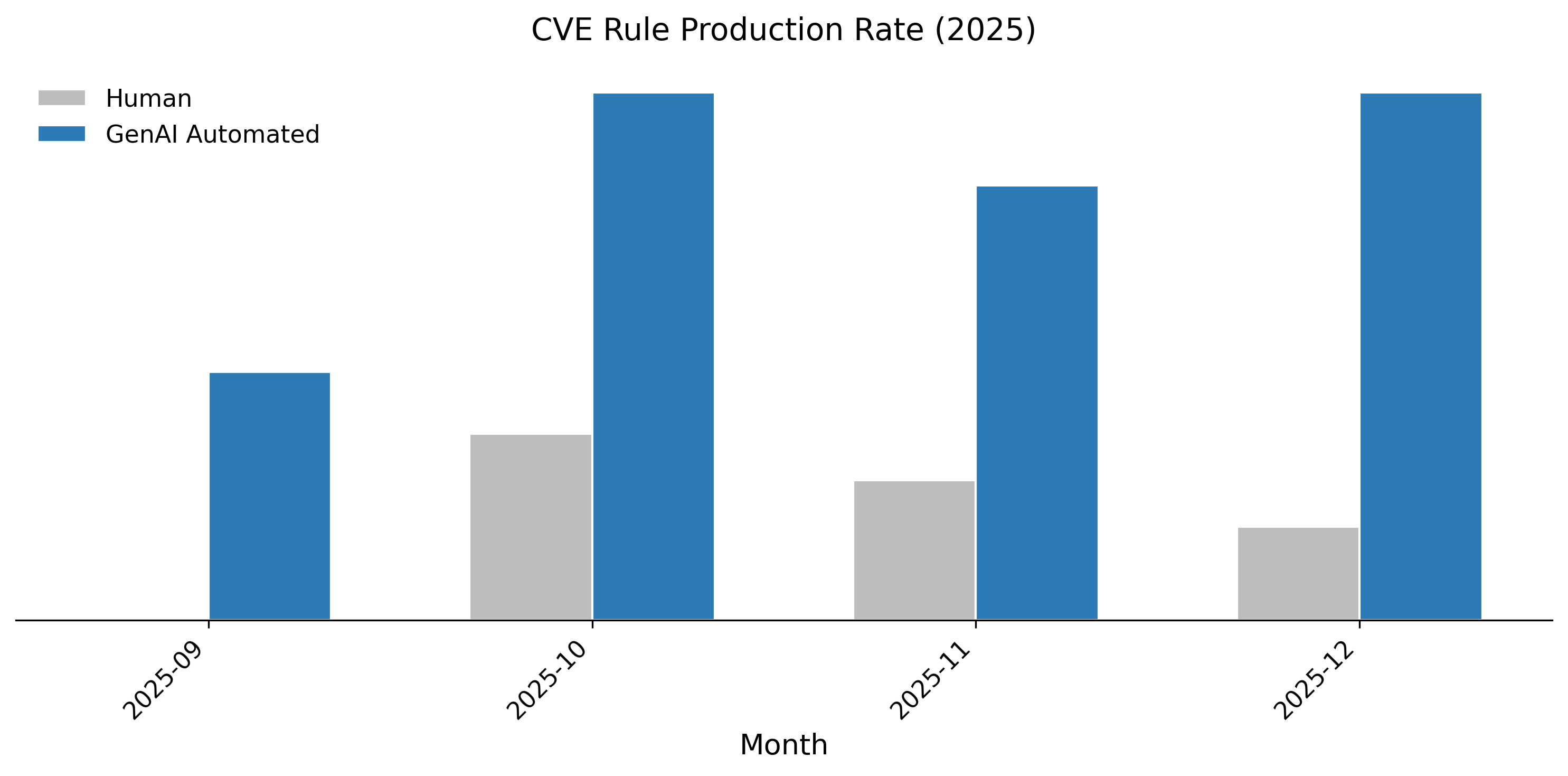}
\caption{CVE Rule Production Rate comparing GenAI Automated versus Human generation in the last four months of 2025, showing productivity improvements with GenAI automation. Absolute qualitative metrics omitted on purpose, see note in the main text.}
\label{fig:production_rate}
\end{figure}

The operational evaluation period spans the final four months of 2025, demonstrating significant improvements in automated CVE rule generation efficiency and productivity. Figure~\ref{fig:production_rate} illustrates the comparative production rates between human-generated and \ruleforge-automated (AI generation) CVE rules during the evaluation period. The data reveals distinct performance characteristics demonstrating the system's operational effectiveness.

The automated system achieved a 336\% productivity advantage compared to human generation during this period. This represents a 419\% increase in AI-generated output relative to earlier baseline performance, while human generation increased by only 38\%. \textit{Relative metrics are reported; absolute counts are omitted to avoid
revealing detection coverage.} Notable surges in automated generation occurred during the evaluation period, demonstrating the system's ability to scale effectively. This productivity improvement validates the effectiveness of our 5×5 generation strategy and systematic feedback integration, confirming the system's capacity to progressively improve rule generation efficiency while maintaining quality standards through the multi-stage validation pipeline.

The volume of CVEs processed for rule generation has been intentionally limited to prevent overwhelming human reviewers who perform the final validation stage. Despite this operational constraint, AI generation (\ruleforge) demonstrated a substantial productivity advantage compared to human generation during the evaluation period. This performance validates the system's capacity to progressively improve rule generation efficiency while maintaining quality standards through the multi-stage validation pipeline.

\section{RULEFORGE EXTENSIONS: EXPLORING GENERALIZATION AND ADAPTABILITY}\label{sec:extensions}

While the LLM-as-a-judge confidence validation system has been successfully integrated into the production \ruleforge~pipeline, two additional research directions were explored to demonstrate the system's potential for generalization and adaptability beyond its current operational scope. These extensions—generating rules from unstructured data and implementing an agentic workflow—are not yet part of the production system but showcase avenues for future development and illustrate \ruleforge's capacity to evolve beyond its initial design constraints.

\subsection{Generating Rules from Unstructured Data}

The production \ruleforge~system relies on structured Nuclei templates, which provide two essential fields: a concise vulnerability description and example HTTP requests demonstrating malicious queries. However, this dependency limits the system's scope, as only a small fraction of documented vulnerabilities are converted to Nuclei templates. To explore whether \ruleforge~could operate on more diverse data sources, we adapted the system to process unstructured vulnerability descriptions from sources such as blog posts, GitHub advisories, and NVD entries.

Unstructured data presents several challenges compared to Nuclei templates. First, vulnerability descriptions often lack concrete examples of malicious HTTP requests, requiring the system to infer attack patterns from textual descriptions alone. Second, unstructured sources frequently contain extraneous information—artifacts from web scraping, navigation elements, links to further reading, and editorial comments—that must be filtered from the relevant vulnerability details. Third, some sources may provide insufficient information or present critical details in formats inaccessible to the LLM, such as screenshots or diagrams.

To address these challenges, we modified the \ruleforge~codebase to accept vulnerability descriptions without example HTTP requests and adjusted the prompts to be more tolerant of longer, less precisely structured text. Critically, we added an explicit option for the system to refuse rule generation when insufficient information is available, preventing the LLM from hallucinating rules based on incomplete data. Testing with intentionally incomplete inputs confirmed that the system appropriately exercises this refusal option rather than generating invalid rules.

We evaluated this approach using a proof-of-concept dataset of ten vulnerabilities obtained from diverse sources: eight from NVD, one from a GitHub advisory, and one from a security blog post. The LLM generated rules for eight of these vulnerabilities. Human evaluation classified four of the generated rules as suitable (usable with minor modifications or acceptable with low confidence) and four as unsuitable, yielding a 40\% overall success rate and a 50\% success rate among generated rules. The small sample size limits the generalizability of these results, but the findings suggest that unstructured data generation is feasible and could expand \ruleforge's applicability.

Analysis of the failure modes revealed that the most common reason for rule rejection was fundamental unsuitability of the vulnerability for detection—cases where no HTTP-based detection rule could effectively mitigate the threat. This suggests that future work should focus on enhancing the system's ability to recognize and refuse such cases earlier in the generation process.

\subsection{Agentic Workflow for Multi-Event-Type Rule Generation}

The production \ruleforge~system currently generates rules exclusively for HTTP events. However, the event classification pipeline supports four event types: HttpEvent, ProcessEvent, DnsEvent, and CloudTrailEvent. Extending \ruleforge~to handle multiple event types would require determining which type of rule is appropriate for each vulnerability—a task well-suited to an agentic workflow architecture.

\subsubsection{ReAct Agentic Framework}

We implemented a proof-of-concept using the DSPy implementation of the ReAct~\cite{yao2022react} agentic workflow framework, in which a general-purpose LLM agent receives vulnerability descriptions and iteratively calls specialized tools to generate appropriate detection rules. The agent reasons about tool outputs to determine whether to make additional tool calls or return a final result. This architecture offers several advantages: it can route vulnerabilities to the appropriate rule generator based on the vulnerability's characteristics, handle cases where rule generation fails by escalating to human review, and gracefully manage situations where no vulnerability is present in the input data.

An important design feature that enables this workflow is \ruleforge's practice of returning detailed reasoning when rule generation fails. When given a vulnerability requiring a different event type, the system not only refuses to generate an HTTP rule but also explains why—for example, stating "This is not describing a web vulnerability or HTTP-based attack." These informative error messages provide the agent with actionable feedback for determining its next steps.

\subsubsection{Tool Architecture}

For the proof-of-concept, we defined three categories of tools:

\begin{itemize}
\item \textbf{Rule generators}: Specialized generators for HTTP rules, process rules, and cloud rules, each tailored to detect vulnerabilities in their respective event types
\item \textbf{Human escalation}: A tool to route complex or ambiguous vulnerabilities to security engineers for manual review
\item \textbf{Abandonment}: A tool allowing the system to terminate processing when the input does not contain a valid vulnerability description
\end{itemize}

\subsubsection{Proof-of-Concept Results}

To validate the agentic workflow concept, we created mock versions of all tools that simulate \ruleforge's actual behavior patterns, including realistic success and failure modes. Testing with diverse inputs demonstrated that the agent behaves as intended:

\begin{itemize}
\item Given a straightforward HTTP vulnerability, the agent made a single call to the HTTP rule generator and returned the result
\item Given a document containing no vulnerability, the agent correctly detected the absence of actionable content and terminated processing
\item Given a vulnerability where the initial rule generator failed, the agent attempted alternative generators and, upon continued failure, escalated the case to human review
\end{itemize}

While the tools are mocked, they accurately replicate \ruleforge's actual outputs, including both successful rule generation and informative error messages. This black-box approach requires no fine-tuning of the agent and demonstrates flexibility: adding support for a new event type (such as DnsEvent) would require only adding the corresponding tool to the agent's available toolkit and including a brief description in the agent's signature.

\subsection{Implications for \ruleforge~Evolution}

These extensions demonstrate that \ruleforge's core architecture can be generalized beyond its initial design constraints. The unstructured data generation capability could expand the system's coverage by enabling rule creation from the broader vulnerability disclosure ecosystem, while the agentic workflow architecture provides a scalable path toward comprehensive automated vulnerability mitigation across all event types supported by the event classification pipeline.

\section{Lessons learned}\label{sec:lessons_learned}

\subsection{LLMs are Overconfident for This Application}

One approach we attempted was to ask more generic questions: for instance, adding a field in the generation to ask the model to specify its \textit{confidence} in its own answer as part of the rule generation process. The LLM is very consistent when asked as part of the rule generation to predict its confidence (its lowest confidence in its own answer is 0.7, its highest is 0.9). It does not appear to demonstrate useful discriminative ability. This aligns with previous work suggesting that LLMs do not have good calibration on security topics~\cite{bhusal2024secure}.


\begin{figure*}[htbp]
\centering
\begin{subfigure}[b]{0.48\textwidth}
\centering
\includegraphics[width=\textwidth]{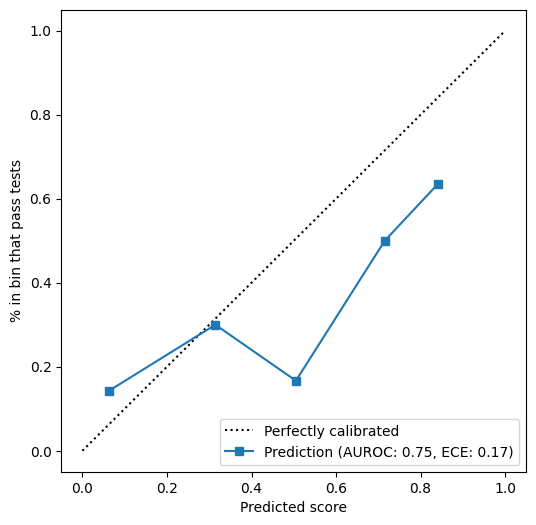}
\caption{Negative phrasing}
\label{fig:negative_phrasing}
\end{subfigure}
\hfill
\begin{subfigure}[b]{0.48\textwidth}
\centering
\includegraphics[width=\textwidth]{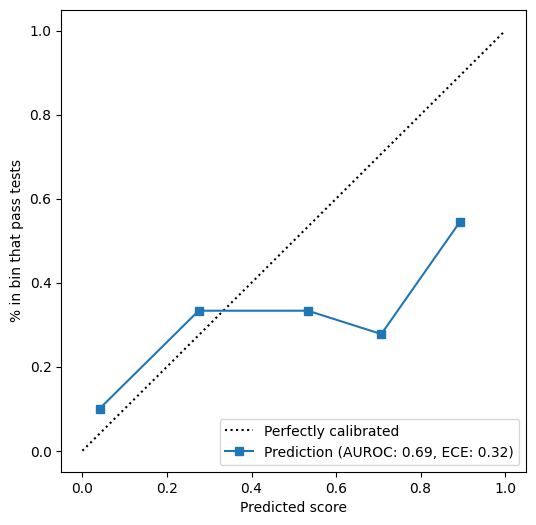}
\caption{Positive phrasing}
\label{fig:positive_phrasing}
\end{subfigure}
\caption{Comparison of LLM-as-a-judge performance using (a) negative phrasing (asking about problems with the rule) versus (b) positive phrasing (asking about rule correctness). Negative phrasing achieves better AUROC and lower ECE, i.e. improves calibration and discriminative power.}
\label{fig:phrasing_comparison}
\end{figure*}

This also manifests in the LLM being more accurate when asked to predict the likelihood that there are problems with the rule, rather than the probability that there are not problems with the rule. Given these results in combination with evidence from literature that LLMs tend to be overconfident~\cite{xiong2023can} and sycophantic~\cite{sharma2023towards}, a good practice for evaluation prompt design is to have the LLM explicitly act as a critic to find flaws, not to ask whether something is correct.

\subsection{Specificity Improves Performance for LLM-as-a-Judge}

We designed the questions with the intention of mimicking human thought processes when evaluating rules. We found that this consultation with experts had an effect on the success of LLM-as-a-judge. More generic questions failed to provide useful estimates, Figure~\ref{fig:phrasing_comparison}. These findings highlight that consulting with experts provides noticeable benefits to the quality of outputs; human experience and expertise is useful when designing prompts.

\subsection{Multiple Candidate Generation Enhances Rule Quality}

Our experience with \ruleforge~demonstrates that generating multiple rule candidates simultaneously improves the likelihood of producing high-quality detection rules. By creating five parallel candidates with different configurations, the system explores diverse detection approaches and increases the chances of generating at least one effective rule per CVE. This parallel generation strategy, combined with iterative refinement based on validation feedback, consistently outperforms single-candidate approaches and contributes substantially to the system's overall success rate in automated rule creation.

\begin{figure*}[htbp]
\centering
\begin{subfigure}{0.48\textwidth}
\centering
\includegraphics[width=\textwidth]{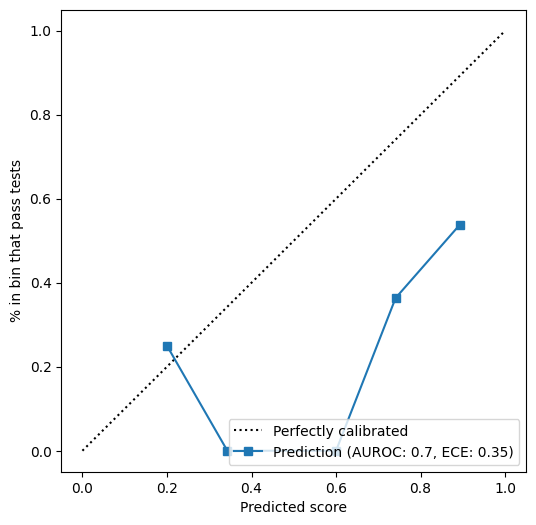}
\caption{Generic confidence questions}
\label{fig:generic_prompt}
\end{subfigure}
\hfill
\begin{subfigure}{0.48\textwidth}
\centering
\includegraphics[width=\textwidth]{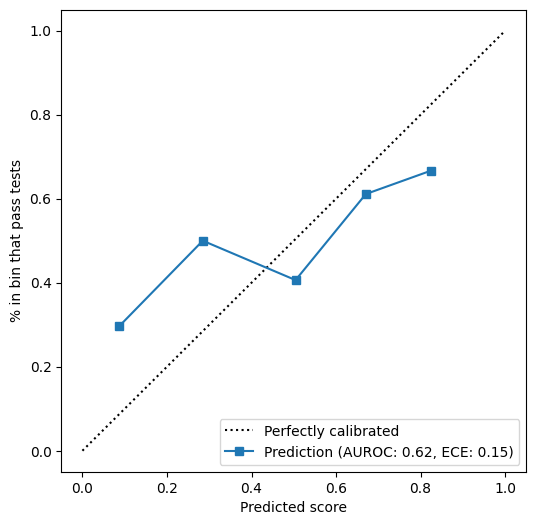}
\caption{Generic FP/FN questions}
\label{fig:generic_fp_fn}
\end{subfigure}
\caption{Performance comparison of generic prompts: (a) generic confidence questions showing poor calibration and (b) generic false positive/negative questions demonstrating worse AUROC than specifically worded prompts.}
\label{fig:generic_performance}
\end{figure*}

\section{Related work}\label{sec:related_work}

Automated vulnerability detection and mitigation has been explored through various approaches, including static analysis, machine learning-based classification, and more recently, large language model-based code generation.

\subsection{LLMs for cybersecurity}

The increased capability of LLMs for various downstream tasks, along with the growing number of reported vulnerabilities in modern software \cite{Ghimire_2025, zemicheal2024llm}, has led to interest in applying LLMs to cybersecurity problems. Benchmarks such as SECURE \cite{bhusal2024secure}, CyberBench \cite{liu2024cyberbench}, and CyberMetric \cite{tihanyi2024cybermetric} aim to quantify the knowledge pretrained LLMs have regarding cybersecurity. The results on these benchmarks suggest that LLMs, particularly proprietary closed-source LLMs, have robust cybersecurity knowledge. LLM systems have been constructed for a variety of cybersecurity applications such as chatbots for cybersecurity \cite{su151713178}, penetration testing\cite{10.1145/3611643.3613083}, vulnerability management \cite{zemicheal2024llm}, and many others.

\subsection{Vulnerability detection}

One area of interest for cybersecurity is vulnerability detection, generally applied to existing systems to detect whether there are vulnerabilities that could be exploited. This has been explored in several arenas, such as C/C++ software repositories \cite{Ghimire_2025} and Docker containers\cite{zemicheal2024llm}. In this area, pretrained LLMs can show competitive results to specifically fine-tuned models \cite{Ghimire_2025}. However, one potential issue with automated vulnerability detection is a high number of false positives, which may decrease trust in the system. Less commonly, LLMs have been studied for vulnerability mitigation in addition
to detection: for instance, \cite{10.1145/3314058.3318167} propose a system
for automatically patching software based on CVEs, and
\cite{Ghimire_2025} suggest mitigation strategies after analysis of CVEs.
More closely related to our work, FALCON~\cite{mitra2025falcon} uses an
agentic LLM framework to autonomously generate IDS rules (Snort and YARA) from
cyber threat intelligence, employing multi-phase validation to achieve 95\%
accuracy. \ruleforge~differs in its focus on HTTP-based detection rules for web
vulnerabilities, its LLM-as-a-judge confidence scoring approach, and its
integration into a production deployment pipeline with feedback-driven
refinement.

\section{CONCLUSION}\label{sec:conclusion}

This paper presents \ruleforge, an AWS internal system that addresses the critical gap between vulnerability disclosure and detection rule creation. As vulnerability disclosure rates continue to outpace manual mitigation capabilities, automated approaches are essential for maintaining security at scale.

Our primary contribution is a feedback-driven refinement mechanism that leverages LLM-as-a-judge confidence validation to iteratively improve detection rule quality. By evaluating rules across sensitivity and specificity dimensions and feeding validation results back for up to five refinement attempts per candidate, the system achieves AUROC of 0.75 and has reduced false positives by 67\% while decreasing rules with no IP matches by 71\% since production deployment on August 1, 2025. This approach demonstrates that confidence-based self-evaluation can effectively guide iterative rule generation when properly constrained.

Additionally, we demonstrate the effectiveness of multiple candidate generation combined with comprehensive five-stage validation including feedback-based regeneration. This approach, which generates five parallel rule candidates per CVE and subjects them to synthetic testing, confidence scoring, IP validation, IP enrichment analysis, and human review increases the probability of generating high-quality detection rules while maintaining production-ready standards through systematic feedback integration.

We present promising directions for future work on \ruleforge's extensibility through rule generation from unstructured data sources (40\% success rate) and an agentic workflow for multi-event-type vulnerability mitigation, illustrating pathways toward fully automated CVE response systems.

Our work offers key insights for the cybersecurity research community: LLM-as-a-judge approaches can provide reliable quality signals for security artifacts when designed with domain expertise, production deployment requires multi-stage validation balancing automation with human oversight, and the vulnerability disclosure-to-mitigation gap represents an opportunity for AI-assisted automation. \ruleforge~demonstrates that generative AI can augment human security expertise at production scale while maintaining the precision requirements critical for security applications.

\subsection{Future Work}

The extensions explored in this work open future work for advancing \ruleforge's capabilities and impact. Refining the unstructured data generation approach to better recognize unsuitable vulnerabilities earlier in the process would enable the system to efficiently expand its coverage across the broader vulnerability disclosure ecosystem, potentially increasing rule generation capacity. Extending \ruleforge~to support Process-Event, DNS-Event, and CloudTrail-Event types represents an opportunity to create a multi-modal vulnerability mitigation platform that addresses the full spectrum of security threats. Successfully adapting the system to these additional event types would demonstrate the generalizability of the LLM-as-a-judge validation approach and the agentic workflow architecture across diverse security domains. Finally, thoughtful production deployment of these extensions—with attention to data quality, tuned failure rates, and effective prioritization mechanisms—promises to amplify security engineer productivity while maintaining the high-quality standards established by the current system. These research directions collectively position \ruleforge~to evolve from a specialized HTTP rule generator into a comprehensive platform for automated vulnerability detection and mitigation at scale.

\begin{acks}
This work was conducted while the first author was full-time employee at AWS and the second author held an internship position at AWS.

We thank the Security Insights team at AWS for their support and feedback throughout this project.
\end{acks}

\bibliographystyle{ACM-Reference-Format}
\bibliography{references}

@misc{zemicheal2024llm,
  title={LLM agents for vulnerability identification and verification of CVEs},
  author={ZeMicheal, Tadesse and Chen, Hsin and Davis, Shawn and Allen, Rachel and Demoret, Michael and Song, Ashley},
  year={2024},
  url={https://ceur-ws.org/Vol-3920/paper09.pdf}
}

@inproceedings{liu2024cyberbench,
  title={Cyberbench: A multi-task benchmark for evaluating large language models in cybersecurity},
  author={Liu, Zefang and Shi, Jialei and Buford, John F},
  booktitle={AAAI 2024 Workshop on Artificial Intelligence for Cyber Security},
  year={2024}
}

@inproceedings{tihanyi2024cybermetric,
  title={CyberMetric: a benchmark dataset based on retrieval-augmented generation for evaluating LLMs in cybersecurity knowledge},
  author={Tihanyi, Norbert and Ferrag, Mohamed Amine and Jain, Ridhi and Bisztray, Tamas and Debbah, Merouane},
  booktitle={2024 IEEE International Conference on Cyber Security and Resilience (CSR)},
  pages={296--302},
  year={2024},
  organization={IEEE}
}

@Article{su151713178,
  AUTHOR = {Arora, Amit and Arora, Anshu and McIntyre, John},
  TITLE = {Developing Chatbots for Cyber Security: Assessing Threats through Sentiment Analysis on Social Media},
  JOURNAL = {Sustainability},
  VOLUME = {15},
  YEAR = {2023},
  NUMBER = {17},
  ARTICLE-NUMBER = {13178},
  URL = {https://www.mdpi.com/2071-1050/15/17/13178},
  ISSN = {2071-1050},
  DOI = {10.3390/su151713178}
}

@inproceedings{10.1145/3611643.3613083,
  author = {Happe, Andreas and Cito, J\"{u}rgen},
  title = {Getting pwn'd by AI: Penetration Testing with Large Language Models},
  year = {2023},
  isbn = {9798400703270},
  publisher = {Association for Computing Machinery},
  address = {New York, NY, USA},
  url = {https://doi.org/10.1145/3611643.3613083},
  doi = {10.1145/3611643.3613083},
  booktitle = {Proceedings of the 31st ACM Joint European Software Engineering Conference and Symposium on the Foundations of Software Engineering},
  pages = {2082--2086},
  numpages = {5},
  location = {San Francisco, CA, USA},
  series = {ESEC/FSE 2023}
}

@article{Ghimire_2025,
   title={HWREx: AI-enabled Hardware Weakness and Risk Exploration and Storytelling Framework with LLM-assisted Mitigation Suggestion},
   ISSN={1557-7309},
   url={http://dx.doi.org/10.1145/3737459},
   DOI={10.1145/3737459},
   journal={ACM Transactions on Design Automation of Electronic Systems},
   publisher={Association for Computing Machinery (ACM)},
   author={Ghimire, Sujan and Lin, Yu-Zheng and Mamun, Muntasir and Chowdhury, Muhtasim Alam and Alemi, Farhad and Cai, Shuyu and Guo, Jinduo and Zhu, Mingyu and Li, Honghui and Saber Latibari, Banafsheh and Rafatirad, Setareh and Satam, Pratik and Salehi, Soheil},
   year={2025},
   month=may
}

@inproceedings{10.1145/3314058.3318167,
author = {Aghaei, Ehsan and Al-shaer, Ehab},
title = {ThreatZoom: neural network for automated vulnerability mitigation},
year = {2019},
isbn = {9781450371476},
publisher = {Association for Computing Machinery},
address = {New York, NY, USA},
url = {https://doi.org/10.1145/3314058.3318167},
doi = {10.1145/3314058.3318167},
booktitle = {Proceedings of the 6th Annual Symposium on Hot Topics in the Science of Security},
articleno = {24},
numpages = {3},
location = {Nashville, Tennessee, USA},
series = {HotSoS '19}
}

@inproceedings{guo2017calibration,
  title={On calibration of modern neural networks},
  author={Guo, Chuan and Pleiss, Geoff and Sun, Yu and Weinberger, Kilian Q},
  booktitle={International conference on machine learning},
  pages={1321--1330},
  year={2017},
  organization={PMLR}
}

@article{yao2022react,
  title={React: Synergizing reasoning and acting in language models},
  author={Yao, Shunyu and Zhao, Jeffrey and Yu, Dian and Du, Nan and Shafran, Izhak and Narasimhan, Karthik and Cao, Yuan},
  journal={arXiv preprint arXiv:2210.03629},
  year={2022}
}

@inproceedings{bhusal2024secure,
  title={SECURE: Benchmarking Large Language Models for Cybersecurity},
  author={Bhusal, Dipkamal and Islam, Md Tanvir and Park, Youngja and Rastogi, Nidhi},
  booktitle={2024 Annual Computer Security Applications Conference (ACSAC)},
  year={2024},
  organization={IEEE}
}

@article{xiong2023can,
  title={Can llms express their uncertainty? an empirical evaluation of confidence elicitation in llms},
  author={Xiong, Miao and Hu, Zhiyuan and Lu, Xinyang and Li, Yifei and Fu, Jie and He, Junxian and Hooi, Bryan},
  journal={arXiv preprint arXiv:2306.13063},
  year={2023}
}

@article{sharma2023towards,
  title={Towards understanding sycophancy in language models},
  author={Sharma, Mrinank and Tong, Meg and Korbak, Tomasz and Duvenaud, David and Askell, Amanda and Bowman, Samuel R and Cheng, Newton and Durmus, Esin and Hatfield-Dodds, Zac and Johnston, Scott R and others},
  journal={arXiv preprint arXiv:2310.13548},
  year={2023}
}

@misc{Nuclei,
  author = {{ProjectDiscovery team}},
  title = {{Nuclei: Fast and customizable vulnerability scanner based on simple YAML based DSL}},
  howpublished = {https://github.com/projectdiscovery/nuclei},
  publisher = {{ProjectDiscovery}},
  year = {2025},
  note = {2025-11-28},
}

@article{khattab2024dspy,
  title={DSPy: Compiling Declarative Language Model Calls into Self-Improving Pipelines},
  author={Khattab, Omar and Singhvi, Arnav and Maheshwari, Paridhi and Zhang, Zhiyuan and Santhanam, Keshav and Vardhamanan, Sri and Haq, Saiful and Sharma, Ashutosh and Joshi, Thomas T. and Moazam, Hanna and Miller, Heather and Zaharia, Matei and Potts, Christopher},
  journal={The Twelfth International Conference on Learning Representations},
  year={2024}
}

@misc{Ryland2023MadPot,
  author = {Ryland, Mark and Mehr, Nima Sharifi},
  title = {{Meet MadPot, a threat intelligence tool Amazon uses to protect customers from cybercrime}},
  year = {2023},
  month = {sep},
  url = {https://www.aboutamazon.com/news/aws/amazon-madpot-stops-cybersecurity-crime},
  note = {{Accessed on: 2025-12-04}}
}

@article{zheng2023judging,
  title={Judging LLM-as-a-Judge with MT-Bench and Chatbot Arena},
  author={Zheng, Lianmin and Chiang, Wei-Lin and Sheng, Ying and Zhuang, Siyuan and Wu, Zhanghao and Zhuang, Yonghao and Lin, Zi and Li, Zhuohan and Li, Dacheng and Xing, Eric P and Zhang, Hao and Gonzalez, Joseph E and Stoica, Ion},
  journal={Advances in Neural Information Processing Systems},
  volume={36},
  year={2023}
}

@article{mitra2025falcon,
  title={FALCON: Autonomous Cyber Threat Intelligence Mining with LLMs for IDS Rule Generation},
  author={Mitra, Shaswata and Bazarov, Azim and Duclos, Martin and Mittal, Sudip and Piplai, Aritran and Rahman, Md Rayhanur and Zieglar, Edward and Rahimi, Shahram},
  journal={arXiv preprint arXiv:2508.18684},
  year={2025}
}

@article{panickssery2024llm,
  title={LLM Evaluators Recognize and Favor Their Own Generations},
  author={Panickssery, Arjun and Bowman, Samuel R and Feng, Shi},
  journal={arXiv preprint arXiv:2404.13076},
  year={2024}
}

@misc{nvd2025dashboard,
  author = {{National Institute of Standards and Technology}},
  title = {{National Vulnerability Database Dashboard}},
  howpublished = {https://nvd.nist.gov/general/nvd-dashboard},
  year = {2025},
  note = {Accessed: 2025-12-04}
}

\newpage

\appendix

\section{CVE to detection rule example}

This appendix provides a concrete example of how \ruleforge~converts CVE vulnerability descriptions into detection rules.

\textbf{Example: CVE-2023-26256}

CVE-2023-26256~\footnote{NVD: \url{https://nvd.nist.gov/vuln/detail/CVE-2023-26256} | Nuclei: \url{https://github.com/projectdiscovery/nuclei-templates/blob/main/http/cves/2023/CVE-2023-26256.yaml}} is a path traversal vulnerability in STAGIL Navigation for Jira. Attackers can read arbitrary files on the server via the \texttt{fileName} parameter in the \texttt{snjFooterNavigationConfig} endpoint. This vulnerability allows unauthorized access to sensitive system files through directory traversal attacks using sequences like \texttt{../../../}.

\textbf{Generated detection rule:}

\begin{lstlisting}[language=json,basicstyle=\small\ttfamily,breaklines=true,frame=single]
{"conditions": [
  {"var": "query_string",
   "comparison": "contains",
   "constant": "../../"},
  {"var": "path",
   "comparison": "equals",
   "constant":
     "/plugins/servlet/snjFooterNavigationConfig"}
 ],
 "conditions_match": "and"}
\end{lstlisting}

This rule demonstrates \ruleforge's ability to identify the key attack patterns (directory traversal sequences in query parameters) and the specific vulnerable endpoint, creating a precise detection mechanism that can identify exploitation attempts while minimizing false positives on legitimate traffic.

\section{\ruleforge~Confidence Prompts}\label{sec:confidence_prompts}

The LLM-as-a-judge confidence validation system is implemented using the DSPy framework with two signature classes that define the input-output structure for sensitivity and specificity evaluation. In accordance with security best practices, the implementation details presented have been generalized to illustrate the core methodology while omitting security-sensitive components that could compromise the production system's effectiveness.

\subsection{Sensitivity Confidence Signature}

The \texttt{RuleSensitivityConfidence} signature evaluates whether a rule fails to flag malicious requests. It takes as input the CVE vulnerability description and the candidate rule, and outputs a probability score indicating the likelihood that the rule misses malicious requests described in the CVE.

\subsection{Specificity Confidence Signature}

The \texttt{RuleSpecificityConfidence} signature evaluates whether a rule targets correlated features rather than the vulnerability itself. It uses the same inputs as the sensitivity signature but outputs a probability score indicating the likelihood that the rule produces false positives by matching benign traffic patterns.

\subsection{Confidence Module Implementation}

The \texttt{ConfidenceModule} orchestrates the confidence evaluation by invoking both sensitivity and specificity assessments using chain-of-thought reasoning. The module:

\begin{enumerate}
\item Invokes \texttt{dspy.ChainOfThought} for calculating sensitivity and specificity scores.
\item Computes complement scores: \texttt{sensitivity\_score = 1 - sensitivity\_result.score} and \texttt{specificity\_score = 1 - specificity\_result.score}
\item Calculates overall confidence as the product: \texttt{confidence = sensitivity\_score × specificity\_score}
\item Returns a \texttt{dspy.Prediction} object containing:
\begin{itemize}
\item Overall confidence score
\item Individual sensitivity and specificity scores
\item Reasoning chains explaining the model's rationale
\end{itemize}
\end{enumerate}

When scores fall below established thresholds (0.5 for sensitivity and 0.7 for specificity), the reasoning chains are appended to the prompt as feedback for iterative rule refinement. This feedback mechanism enables the system to progressively improve rule quality through multiple generation attempts.

\end{document}